\documentclass[10pt,twocolumn,letterpaper]{article}

\usepackage[pagenumbers]{cvpr} 

\usepackage{amsmath}
\usepackage{algorithm}
\usepackage[noend]{algpseudocode}

\definecolor{cvprblue}{rgb}{0.21,0.49,0.74}
\usepackage[pagebackref,breaklinks,colorlinks,allcolors=cvprblue]{hyperref}
\usepackage[symbol]{footmisc}

\def\paperID{xxxxx} 
\def\confName{CVPR}
\def\confYear{2025}

\title{Scaling Mesh Generation via Compressive Tokenization}

\author{Haohan Weng$^{14}$,~
Zibo Zhao$^{24}$\footnotemark[1],~
Biwen Lei$^4$,~
Xianghui Yang$^4$,~
Jian Liu$^{4}$,~
Zeqiang Lai$^4$,~
Zhuo Chen$^4$ \\
\vspace{2mm}
Yuhong Liu$^4$,~
Jie Jiang$^4$,~
Chunchao Guo$^4$,~
Tong Zhang$^1$,~
Shenghua Gao$^3$,~
C. L. Philip Chen$^1$ \\
$^1$South China University of Technology,
$^2$ShanghaiTech University,
$^3$University of Hong Kong \\
$^4$Tencent Hunyuan\\
\url{https://whaohan.github.io/bpt}
}

\begin{document}

\twocolumn[{%
\maketitle
\begin{center}
\vspace{-4mm}
    \centering
    \captionsetup{type=figure}
    \includegraphics[width=\textwidth]{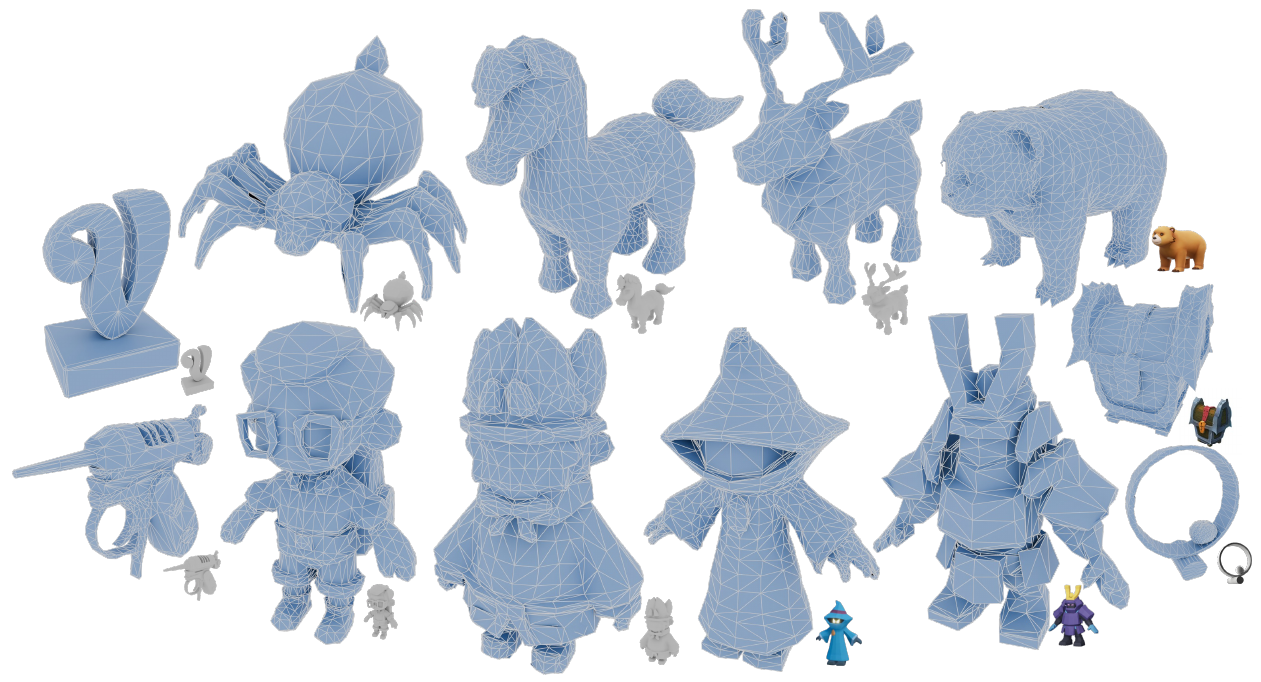}
    \captionof{figure}{\textbf{Generated meshes conditioned on images or point cloud sampled from dense meshes.} Our model can generate meshes up to 8k faces based on the proposed compressive tokenization. The lower right dense meshes or images represent the conditions.}
    \label{fig: demo}
\end{center}%
}]
\footnotetext[1]{Corresponding Author}
\begin{abstract}

We propose a compressive yet effective mesh representation, \textbf{Blocked and Patchified Tokenization (BPT)}, facilitating the generation of meshes exceeding 8k faces. 
BPT compresses mesh sequences by employing block-wise indexing and patch aggregation, reducing their length by approximately 75\% compared to the original sequences. 
This compression milestone unlocks the potential to utilize mesh data with significantly more faces, thereby enhancing detail richness and improving generation robustness.
Empowered with the BPT, we have built a foundation mesh generative model training on scaled mesh data to support flexible control for point clouds and images. 
Our model demonstrates the capability to generate meshes with intricate details and accurate topology, achieving SoTA performance on mesh generation and reaching the level for direct product usage.

\end{abstract}

\begin{figure*}
\centering
\includegraphics[width=\textwidth]{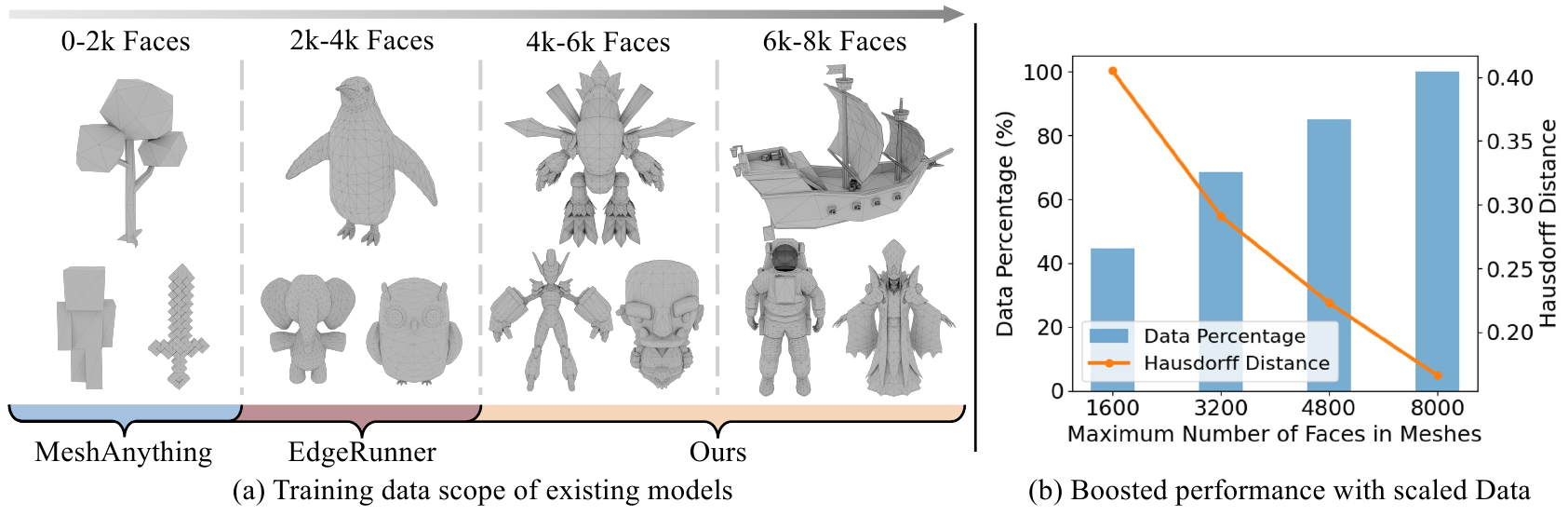}
\caption{\textbf{Scaling data for mesh generation with BPT.} 
(a) Existing models can only handle meshes with at most 4k faces, which still lack intricate details. Empowered by BPT, our model can leverage meshes exceeding 8k faces, effectively extending the training scope for mesh generation.
(b) We train the same model on meshes with different maximum numbers of faces. 
As the number of mesh faces increases, the performance of mesh generation significantly improves, highlighting the value of high-poly training data.
The generation performance is measured by the Hausdorff distance between the input point cloud and generated meshes (a lower distance indicates a better performance).  
}
\vspace{-4mm}
\label{fig: data}
\end{figure*}

\section{Introduction}


Meshes, the cornerstone of 3D geometric representation, are widely utilized in various applications, including video games, cinematic productions, and simulations.
Despite their widespread adoption, the meticulous craft of meshes with functional topologies demands substantial design effort. This labor-intensive process acts as a bottleneck, impeding the evolution of 3D content creation and the progress of immersive human-computer interaction.
Recent research~\cite{siddiqui2023meshgpt,chen2024meshxl,weng2024pivotmesh,chen2024meshanything,chen2024meshanythingv2,tang2024edgerunner} has tried to automate the mesh sculpting process via auto-regressive Transformers.
These methods directly generate vertices and faces as human-crafted to maintain the high-quality mesh topology, yielding promising results on low-poly mesh generation.


The foundation of modeling meshes with auto-regressive transformers is mesh tokenization, which converts the mesh into a one-dimensional sequence. 
PolyGen~\cite{nash2020polygen} and MeshXL~\cite{chen2024meshxl} directly tokenize the mesh by converting the vertex coordinates to sorted triplets, each defining a tuple of quantized 3D coordinates. 
They learn the one-dimensional sequence with a joint or two separate auto-regressive transformers.
MeshGPT~\cite{siddiqui2023meshgpt} and its variants~\cite{weng2024pivotmesh,chen2024meshanything} utilize an auto-encoder to convert meshes into latent sequences. 
MeshAnythingv2~\cite{chen2024meshanythingv2} and Edgerunner~\cite{tang2024edgerunner} propose improved tokenization methods to compress vanilla mesh sequences further. 
However, these methods still convert meshes to relatively long sequences, limiting the ability of generative models to learn with high-poly meshes. 
\textit{To scale up mesh generation, a more compressive representation is demanded to extend the scope of the training data.}

In this paper, we propose a compressive yet efficient mesh representation called Blocked and Patchified Tokenization (BPT).
BPT converts Cartesian coordinates to block-wise indexes, which makes the initial attempt to compress mesh tokens at the vertex level.
Then, we select the vertices connected with most faces (i.e., the highest vertex degree) as the patch center. The faces around the center vertices are aggregated as patches, compressing mesh tokens at the face level.
Our approach significantly reduces the length of the vanilla mesh sequence by around 75\%, achieving the SoTA compression ratio across existing tokenization.
Empowered with BPT, our mesh generative model can utilize millions of meshes with intricate details, significantly improving its performance and robustness. 


BPT facilitates a wide range of 3D applications. We demonstrate its effectiveness via conditional mesh generation on point clouds and images.
Our model empowers even unprofessional users to produce meshes at the product-ready level.
Its applicability spans a spectrum of practical domains of 3D content creation, revealing the dawn of a new era in 3D generation.

Our contributions can be summarized as follows:
\begin{itemize}[leftmargin=*]
    \item We introduce Blocked and Patchified Tokenization (BPT), a compressive yet effective tokenization with a state-of-the-art compression ratio of around 75\%.
    \item Empowered by BPT, we investigated the scaling of mesh data across diverse face configurations, revealing that incorporating extended data improves generation performance and robustness.
    \item We build a mesh foundation model that supports conditional generation based on images and point clouds, enabling users to create product-ready 3D assets.
\end{itemize}

\begin{figure*}[t]
\centering
\includegraphics[width=\textwidth]{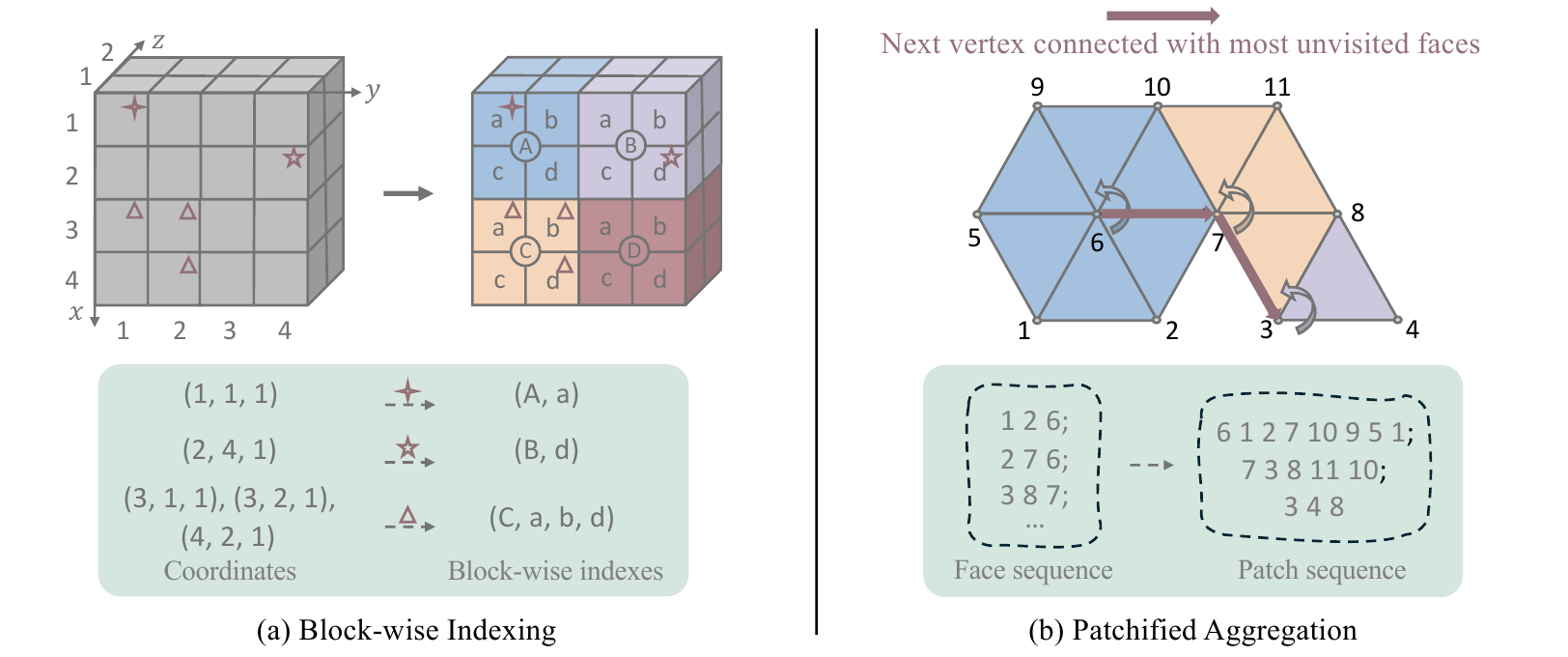}
\caption{\textbf{The proposed Blocked and Patchified Tokenization (BPT).} (a) We convert the coordinates from the Cartesian system to block-wise indexes. The coordinates are first separated equally into several blocks. Then, vertices inside each block are located with 1-dim indexes. (b) The nearby faces are aggregated as patches to compress the mesh sequence. Each patch center is set as the vertex connected with the most unvisited faces. Subsequently, other vertices within the patch are included in the subsequence to create a complete patch.
}
\label{fig: tokenization}
\vspace{-4mm}
\end{figure*}


\section{Scaling Data with Compressive Tokenization}
\label{sec: data}
In this section, we discuss the significant improvement brought by the scaled data, showing the necessity and importance of the proposed compressive tokenization.

\vspace{-4mm}
\paragraph{Extend the training data scope.}
One of the keys to improving current mesh generative models is to extend its training scope (i.e., effectively utilizing the high-poly meshes).
The training data of existing mesh generative models predominantly consists of low-poly meshes, significantly limiting their ability to express intricate details and handle complicated inputs.
As shown in Figure \ref{fig: data}(a), the training data from previous methods still lacks sufficient detail. 
In contrast, BPT accommodates meshes exceeding 8k faces, showing the potential to utilize vast amounts of high-quality mesh data with rich details.

\vspace{-4mm}
\paragraph{Boosted performance and robustness.}
As shown in Figure \ref{fig: data}(b), we train the same model on meshes with different maximum numbers of faces. 
To qualitatively demonstrate the performance improvement, we compute the Hausdorff distance between the input point cloud and generated meshes (the lower distance indicates better performance).
The figure shows the surprising enhancement afforded by the scaled data. 
It illustrates the positive correlation between the number of faces and the performance of generative models. As the mesh data's complexity increases, the generative models' performance and robustness are significantly boosted.

\section{Blocked and Patchified Tokenization}


This section will explore Blocked and Partchified Tokenization (BPT) as illustrated in Figure \ref{fig: tokenization} and Algorithm \ref{alg:patchified_aggregation}, which involves block-wise indexing and patch aggregation. 
We will begin by discussing the vanilla mesh tokenization in Section \ref{sec: preliminary}. 
Following that, Section \ref{sec: indexing} will detail the conversion of vertex representations from Cartesian coordinates to block-wise indexes. 
In Section \ref{sec: patchified aggregation}, we will introduce how to directly aggregate nearby faces into patches and model these patches.
With the implementation of BPT, we can achieve a state-of-the-art compression ratio of approximately 75\% while effectively preserving the mesh topology.

\subsection{Preliminary}
\label{sec: preliminary}

A triangle mesh $\mathcal{M} = (f_1, f_2, ..., f_n)$ with $n$ faces can be formulated as the composition of faces $f_i$.
\begin{equation}
\begin{aligned}
f_i &= (v_{i1}, v_{i2}, v_{i3}) \\
&= (x_{i1}, y_{i1}, z_{i1};~ x_{i2}, y_{i2}, z_{i2};~ x_{i3}, y_{i3}, z_{i3})
\end{aligned}
\end{equation}
where each face $f_i$ consists of 3 vertices and each vertex $v_i$ contains 3D coordinates $(x_i, y_i, z_i)$ discretized with a 7-bit uniform quantization. The vertices are sorted by $z$-$y$-$x$ coordinates from lowest to highest, and the faces are ordered by their lowest vertices.

\begin{table*}[htbp]
\centering
\caption{\textbf{Compression ratio over different mesh tokenization methods.}}
\resizebox{\linewidth}{!}{
\begin{tabular}{cccccccc}
\toprule
Method  &MeshXL \cite{chen2024meshxl} &MeshAnything \cite{chen2024meshanything} &MeshGPT \cite{siddiqui2023meshgpt}  &PivotMesh \cite{weng2024pivotmesh} &MeshAnythingv2 \cite{chen2024meshanythingv2} &EdgeRunner \cite{tang2024edgerunner}  &BPT  \\
\midrule
Compression Ratio $\downarrow$  &1.00  &1.00  &0.67 &0.67 &0.46 &0.47  &\textbf{0.26} \\
\bottomrule
\label{tab: tokenization}
\end{tabular}
}
\vspace{-4mm}
\end{table*}

\subsection{Block-wise Indexing}
\label{sec: indexing}

\paragraph{Naive indexing.} 
The basic idea is to convert a 3D Cartesian coordinate $(x_i,y_i,z_i)$ into a scaler index $I_i$. With the 7-bit quantization setting (resolution $r=128$), the coordinates $x_i, y_i, z_i$ are within the range \([0, r-1]\), thus the indexes $I_i$ are ranged from \([0, r^3-1]\), which leads to an unaffordable vocabulary size $r^3$ for Transformer.

\vspace{-4mm}
\paragraph{Indexing in each block.}
As shown in Figure \ref{fig: tokenization}(a), we equally partition the whole coordinate system into several blocks and index the coordinates as offsets in each block. Specifically, we equally separate the coordinates along each axis into $B$ segments, where the length of each segment is $O$. Thus, for the vertex $v_i = (x_i, y_i, z_i)$, the block-wise indexing $(b_i, o_i)$ can be compute as follow:
\begin{equation}
\begin{aligned}
b_i &= (x_i \mid O)\cdot B^2 + (y_i \mid O)\cdot B + z_i\mid O \\
o_i &= (x_i\% O) \cdot O^2 + (y_i\% O)\cdot O + z_i\% O
\end{aligned}
\label{equ: block-wise}
\end{equation}
where $\mid$ denotes denotes division with no remainder and \% denotes the modulo operation.
The block index $b_i\in[0, B^3-1]$ and offset index $o_i\in[0, O^3-1]$ are all discrete integers, $B^3$ is the number of blocks and $O^3$ is the number of offsets in each block. 

\vspace{-4mm}
\paragraph{Block indexes compression.}
As the coordinates are sorted in the $z$-$y$-$x$ order, the next predicted vertex frequently occurs in the same block.
Thus, we can combine all the adjacent block indexes into one index to save more length. Specifically, for vertices $v_i, v_{i+1}, ..., v_n$ in the same block (i.e., their block index are the same), they can be represented as follows:
\begin{equation}
\begin{aligned}
(v_i, v_{i+1}, ..., v_n) &= (b_i, o_i, b_i, o_{i+1}, ..., b_i, o_n) \\
&= (b_i, o_i, o_{i+1}, ..., o_n)
\end{aligned}
\label{equ: block-compress}
\end{equation}
With the proposed block-wise indexing, a vertex can be represented with at most two tokens. The compression ratio so far is around 50\%.

\subsection{Patchified Aggregation}
\label{sec: patchified aggregation}
In vanilla mesh representation, each vertex appears as many times as the number of its connected faces, which results in considerable redundancy.
Similar to image generation \cite{peeblesScalableDiffusionModels2023}, we propose aggregating the faces connected to the same vertex as a patch without overlapping. 

\vspace{-4mm}
\paragraph{Next Patch Prediction.}

A crucial step in patchifying faces is finding each patch's center vertex.
First, we initialize the whole sorted face sequence as unvisited.
Then we select the first unvisited face and designate the vertex \textit{connected with the most unvisited faces as the patch center}, as shown in Figure~\ref{fig: tokenization}(b).
Consequently, all faces connected to the center vertex are aggregated to form a patch, denoted as $P_c = (v_c, v_1, v_2), (v_c, v_2, v_3), ..., (v_c, v_{n-1}, v_{n})$, we can convert it as follows:
\begin{equation}
    P_c = (v_c, v_1, v_2, ..., v_n)
\label{equ: aggregate}
\end{equation}
where $v_c$ is the patch center and $v_{1:n}$ are other vertices in the patch. Then, we mark all the faces in the patch as visited and find the next unvisited face by sorting order until all the faces are visited.
In this way, the occurrences of the patch center vertex \( v_c \) are reduced (average 6) to just 1. Additionally, for most cases, the occurrences of other vertices \( v_{1:n} \) are decreased from the number of connected faces to the number of connected patches.

\vspace{-4mm}
\paragraph{Seperate patches with dual-block indexes.}

An intuitive solution to denote the termination of each patch is appending a special token. However, this increases the length of the mesh sequence and the training cost.
In contrast, we leverage the dual-block indexes to denote the starting of a patch. 
We create two vocabularies for block indexes, one for center vertex $v_c$ and the other for the common vertices $v_{1:n}$. The offset vocabulary is shared by both the center vertex and other vertices.
\begin{equation}
    P_c = (b'_c, o_c, b_1, o_1, b_2, o_2, ..., b_n, o_n)
\label{equ: dual-block}
\end{equation}

The advantages of patch aggregation are twofold. First, a substantial decrease in vertex repetition leads to a condensed mesh sequence. 
Secondly, the inherent clustering of adjacent faces augments the sequence's spatial locality, thereby simplifying its generation. With Patchified Aggregation, the sequence length yields another 50\% shorter.

\begin{algorithm}[t]
\caption{Blocked and Patchified Tokenization (BPT)}
\label{alg:patchified_aggregation}
\begin{algorithmic}[1]
\Procedure{BPT}{$\mathcal{M}$}
\State $P \gets \emptyset$  \Comment{List of patches}
\While{$\mathcal{M} \neq \emptyset$}
\State $f \gets$ select unvisited face in $\mathcal{M}$
\State $v_c \gets$ vertex in $f$ with most unvisited faces
\State $P_c \gets$ $\texttt{Aggregate}(v_c)$ \Comment{Eq. \ref{equ: aggregate}}
\State $P_c \gets \texttt{Blockwise-Index}(P_c)$ \Comment{Eq. \ref{equ: block-wise} \& \ref{equ: block-compress}}
\State $P_c \gets \texttt{Dual-Block}(P_c)$ \Comment{Eq. \ref{equ: dual-block}}
\State $P \gets P \cup P_c, ~~$$\mathcal{M} \gets \mathcal{M} - P_c$ 
\EndWhile
\State \Return $P$
\EndProcedure
\end{algorithmic}
\end{algorithm}

\begin{figure}[t]
\vspace{-4mm}
\centering
\includegraphics[width=0.45\textwidth]{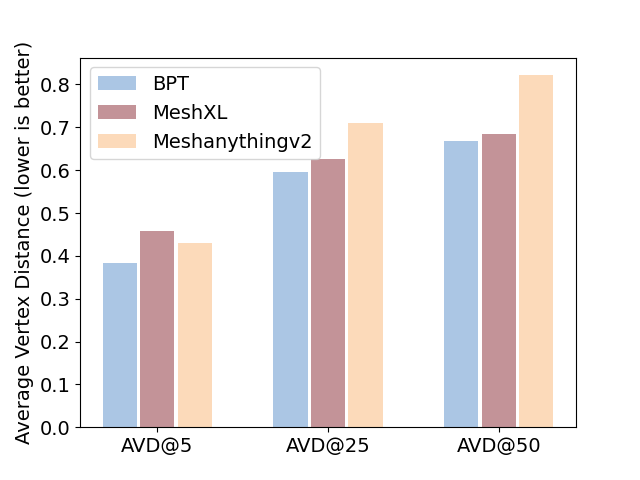}
\caption{\textbf{The average vertex distance with previous $t$ vertices} (denoted as AVD@t, lower is better). Among various context lengths $t$, BPT achieves the lowest AVD, showing its locality for effective mesh modeling.
}
\label{fig: locality}
\vspace{-4mm}
\end{figure}

\subsection{Properties of BPT}

\paragraph{Compactness: free lunch for training and inference efficiency.}
BPT compresses mesh from two perspectives. For the vertex level, one vertex is represented with at most two tokens (one for inner-block indexing and two for inter-block indexing). 
At the face level, faces are aggregated into patches to reduce the repeated vertices that are shared with adjacent faces.
As shown in Table \ref{tab: tokenization}, the proposed BPT achieves the state-of-the-art compression ratio of the mesh sequence.
Consequently, \textit{BPT earns free training and inference efficiency for all mesh generative models.}

\paragraph{Locality: effective mesh sequence modeling.}
The locality is essential for avoiding long-range dependency between tokens in the mesh sequence for transformer modeling.
In BPT, faces are aggregated into patches for generation, guaranteeing each patch's locality.
We quantize the locality with the Average Vertex Distance between vertices and their previous $t$ vertices (denoted as AVD@$t$), as shown in Figure \ref{fig: locality}.
For different values of $t$, BPT achieves the best locality.
Furthermore, as the visual comparison shown in Section \ref{sec: pc-cond}, models based on Adjacent Mesh Tokenization (AMT) \cite{chen2024meshanythingv2} produce meshes with more incomplete meshes. While benefiting from the vital locality, models based on BPT avoid learning the long-range dependency in mesh sequence, thus fundamentally alleviating such problems.

\section{Mesh Generation based on BPT}

The proposed tokenization can be easily applied to mesh generation with high performance and robustness. This section details the architecture of conditional mesh generation on different modalities, including point clouds and images. We use a standard auto-regressive Transformer with parameters $\theta$ to model the sequence with our tokenization and leverage cross-attention for conditioning.
The token sequences are modeled with a standard auto-regressive Transformer with parameter $\theta$, maximizing the log probability. The \textit{cross-attention} is leveraged for various conditions $c$.
\begin{equation}
    L(\theta) =  \prod_{i=1}^{|P|} p(p_{i} | p_{1:i-1}, c;\theta),
\end{equation}
\paragraph{Point-cloud to Mesh.}
We leverage a pre-trained point cloud encoder in Michelangelo-like \cite{zhao2024michelangelo} architecture with larger parameters. 
Furthermore, we add the linear projection and layer-norm layer before cross-attention for a more stable conditioning injection. To yield better generalization ability, we sample 50k points from the mesh and randomly select 4096 points as the condition for each iteration. 

\paragraph{Image to Mesh.}

With the trained point-cloud conditional generative model, we further align the image latent to the point-cloud latent with an additional diffusion model as Michelangelo \cite{zhao2024michelangelo}.
Specifically, we leverage DiT architecture \cite{peeblesScalableDiffusionModels2023} to generate the point-cloud condition produced by the point-cloud encoder.
Our diffusion model is conditioned on the image features extracted from the pre-trained DINO encoder \cite{oquab2023dinov2}.
By feeding the generated point cloud feature into the trained point-cloud conditional generative model, our model can produce meshes faithfully following the given images.


\begin{figure*}[htbp]
\centering
\includegraphics[width=\textwidth]{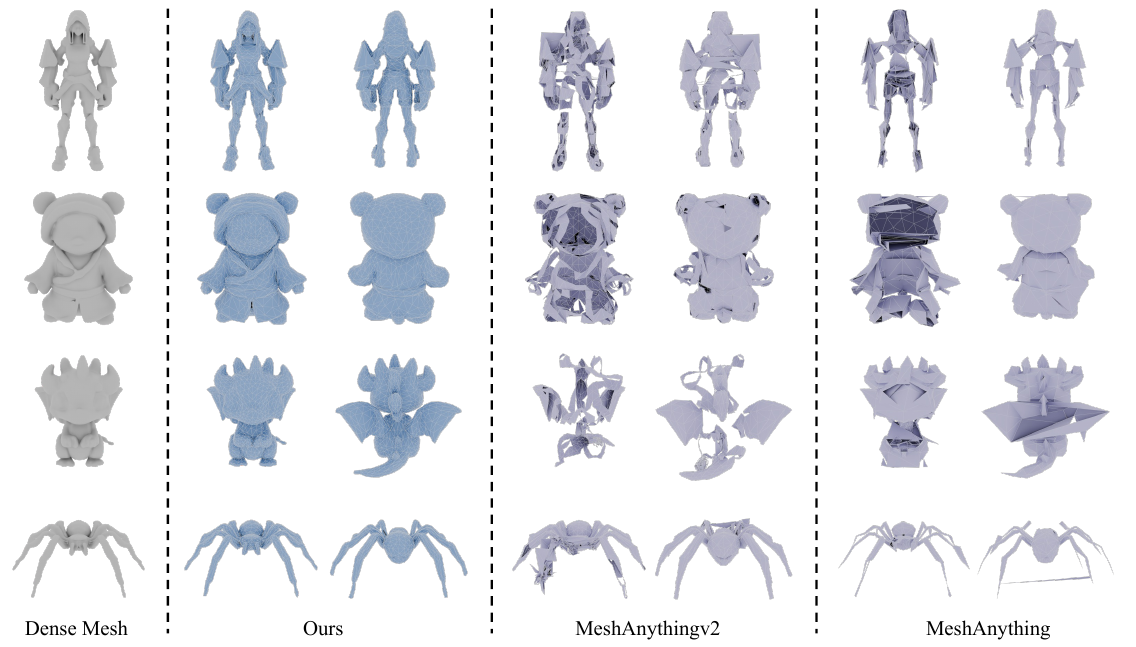}
\caption{\textbf{Comparision on point-cloud conditional generation.} All the meshes are generated conditioned on the point cloud sampled from dense meshes. Our model can recover the details of dense meshes while maintaining high-quality topology.
}
\vspace{-4mm}
\label{fig: pc_cond}
\end{figure*}

\section{Experiments}

\subsection{Experiment Settings}

\paragraph{Datasets.}

Our model is trained on the mixture of ShapeNetV2 \citep{chang2015shapenet}, 3D-FUTURE \cite{fu20213d}, Objaverse \citep{deitke2023objaverse}, Objaverse-XL \citep{deitke2024objaverse} and internal data licensed from 3D content providers.
We manually filter the objects with poor topology (e.g., scanning data) without decimation. 
We further use an effective filtering strategy to maximize data utilization, which filters meshes via their tokenized length based on the transformer's context window 9600.
To trade off the generalizability and mesh quality, we first trained the model on large-scale data with around 1.5M meshes and then further fine-tuned it on 0.3M high-quality meshes.

\paragraph{Baselines.}

We benchmark our approach against leading mesh generation methods: \textbf{MeshAnything} \citep{chen2024meshanything}, which introduces point-cloud conditioned mesh generation model with a noise-resistant decoder to boost mesh generation quality;
\textbf{Meshanythingv2} \citep{chen2024meshanythingv2}, which proposes a new tokenization method called Adjacent Mesh Tokenization (AMT) to achieve half the token sequence length on average compared with vanilla representation;
\textbf{EdgeRunner} \citep{tang2024edgerunner}, which features a novel mesh tokenization algorithm for efficient compression and a fixed-length latent space for better generalization. Since there is no publicly accessible code of Edgerunner, we can only conduct qualitative experiments with EdgeRunner on cases demonstrated on its webpage.

\paragraph{Metrics.}

For point cloud conditional generation, we apply the following metrics: 1) \textbf{Chamfer Distance(CD)}: It calculates the minimum cost required to transform one point set into another by summing the shortest distances from each point in one set to the nearest point in the other set. 2) \textbf{Hausdorff Distance(HD)}: It quantifies the maximum discrepancy between two point sets by measuring the greatest distance from any point in one set to its nearest neighbor in the other set.
For both CD and HD, a lower distance indicates a better performance.
We compute these metrics on 1024-dim point clouds uniformly sampled from meshes.
For image conditional generation, we only conducted qualitative comparisons due to the lack of comparison samples for Edgerunner.

\vspace{-4mm}
\paragraph{Implementation Details.}

Our auto-regressive Transformer has 24 layers with a hidden size of 1024.
It is trained on 4 8×L40 machines for around 7 days with batch size 2 for each GPU.
The temperature used for sampling is set to 0.7 to balance the quality and diversity.
We use flash attention for all Transformer architecture and bf16 mixed precision to speed up the training process. 
We use AdamW \cite{loshchilov2017decoupled} as the optimizer with $\beta_1 = 0.9$ and $\beta_2=0.99$ with a learning rate of $10^{-4}$ for all the experiments.

\begin{table}[t]
\centering
\caption{\textbf{Quantitative comparison with baselines.} With the proposed BPT, our model can utilize meshes with many more faces, thus greatly improving the generation performance and robustness.}
\resizebox{\linewidth}{!}{
\begin{tabular}{ccc}
\toprule
Method &Hausdorff Distance~$\downarrow$  &Chamfer Distance~$\downarrow$ \\
\midrule
MeshAnything \cite{chen2024meshanything} &0.301   &0.136 \\
MeshAnythingv2 \cite{chen2024meshanythingv2} &0.265   &0.114 \\
Ours  &\textbf{0.166} &\textbf{0.094}  \\
\bottomrule
\label{tab: comparison}
\vspace{-4mm}
\end{tabular}
}
\end{table}

\subsection{Point-cloud to Mesh Generation}
\label{sec: pc-cond}

To benchmark the capability of point-cloud conditional generation, we use around 500 dense meshes generated from other models for testing.
We select MeshAnything \cite{chen2024meshanything}, and MeshAnythingV2 \cite{chen2024meshanythingv2} as the baselines. 
We calculate the Hausdorff distance and Chamfer distance between the point clouds sampled from the dense meshes and generated meshes.
The quantitative comparisons are shown in Table \ref{tab: comparison}, indicating our model achieves state-of-the-art performance for both Hausdorff distance and Chamfer distance with significant improvement.
We also qualitatively compare Meshanything and MeshanythingV2 as shown in Figure \ref{fig: pc_cond}. 
Limited with inefficient representations, MeshAnything and MeshAnythingV2 are only trained on meshes with maximum faces of 800 and 1600 separately.
Thus, these models struggle to generate meshes with intricate details and produce more incomplete meshes, as shown in the qualitative comparison.
In contrast, with the proposed BPT, our model can utilize meshes with many more faces and learn the mesh sequence with vital locality, thus significantly improving the generation performance. 

\begin{figure*}[htbp]
\centering
\includegraphics[width=\textwidth]{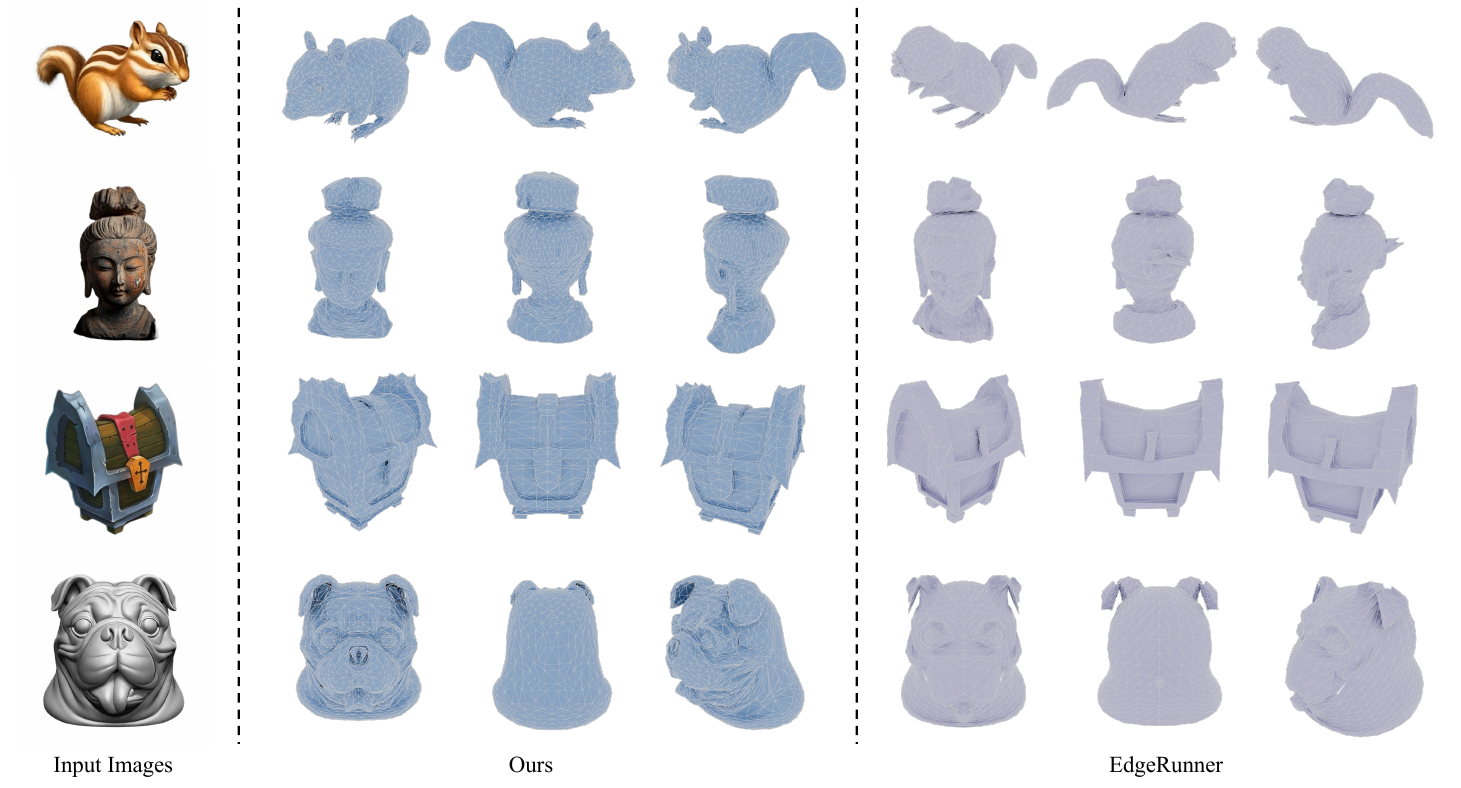}
\caption{\textbf{Comparision on image conditional generation.} The results of Edgerunner are downloaded from its gallery for faithful and comprehensive evaluations. The meshes generated from our model have more details and faithfully follow the input image.
}
\vspace{-2mm}
\label{fig: img_cond}
\end{figure*}

\subsection{Image to Mesh Generation}


There are minimal works \cite{chen2024meshxl,tang2024edgerunner} to generate meshes conditioned on images directly, and none of them are publicly accessible. 
To demonstrate the proposed BPT's effectiveness, we download Edgerunner's gallery from its webpage and generate the meshes with our model conditioned on the same images.
With BPT, our model is trained on meshes with more faces, and thus, it can produce meshes with more details.
As shown in Figure \ref{fig: img_cond}, the meshes generated from our model faithfully follow the input images in more detail compared with Edgerunner's.

\begin{table}[t]
\centering
\caption{\textbf{Abaltion study on block size and offset size.} The block size $8$ and offset size $16$ achieve the best generation performance.}
\resizebox{0.9\linewidth}{!}{
\begin{tabular}{ccc}
\toprule
($|B|$, $|O|$)  & Hausdorff Distance~$\downarrow$ &Chamfer Distance~$\downarrow$ \\
\midrule
(4, 32)  &0.209  &0.111  \\
(8, 16)  &\textbf{0.166} &\textbf{0.094} \\
(16, 8)  &0.256  &0.126 \\
\bottomrule
\label{tab: ablation}
\end{tabular}
}
\vspace{-8mm}
\end{table}

\begin{figure}[htbp]
\centering
\includegraphics[width=0.9\linewidth]{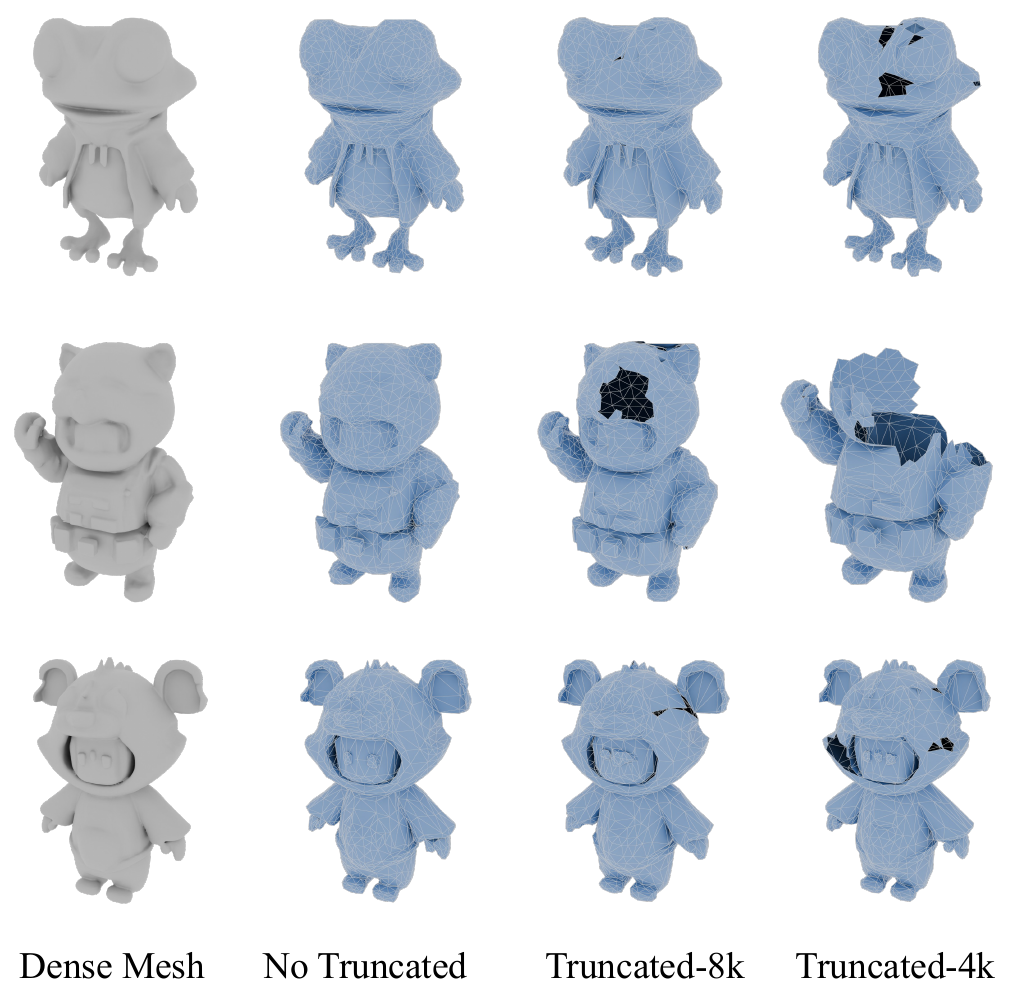}
\caption{\textbf{Trained on truncated meshes with the different context windows, while inference with sliding window attention.} 
Our results show that such truncation would reduce the model's robustness and make it easy to produce incomplete meshes.
}
\vspace{-4mm}
\label{fig: sw_ablation}
\end{figure}

\begin{figure*}[t]
\centering
\includegraphics[width=\textwidth]{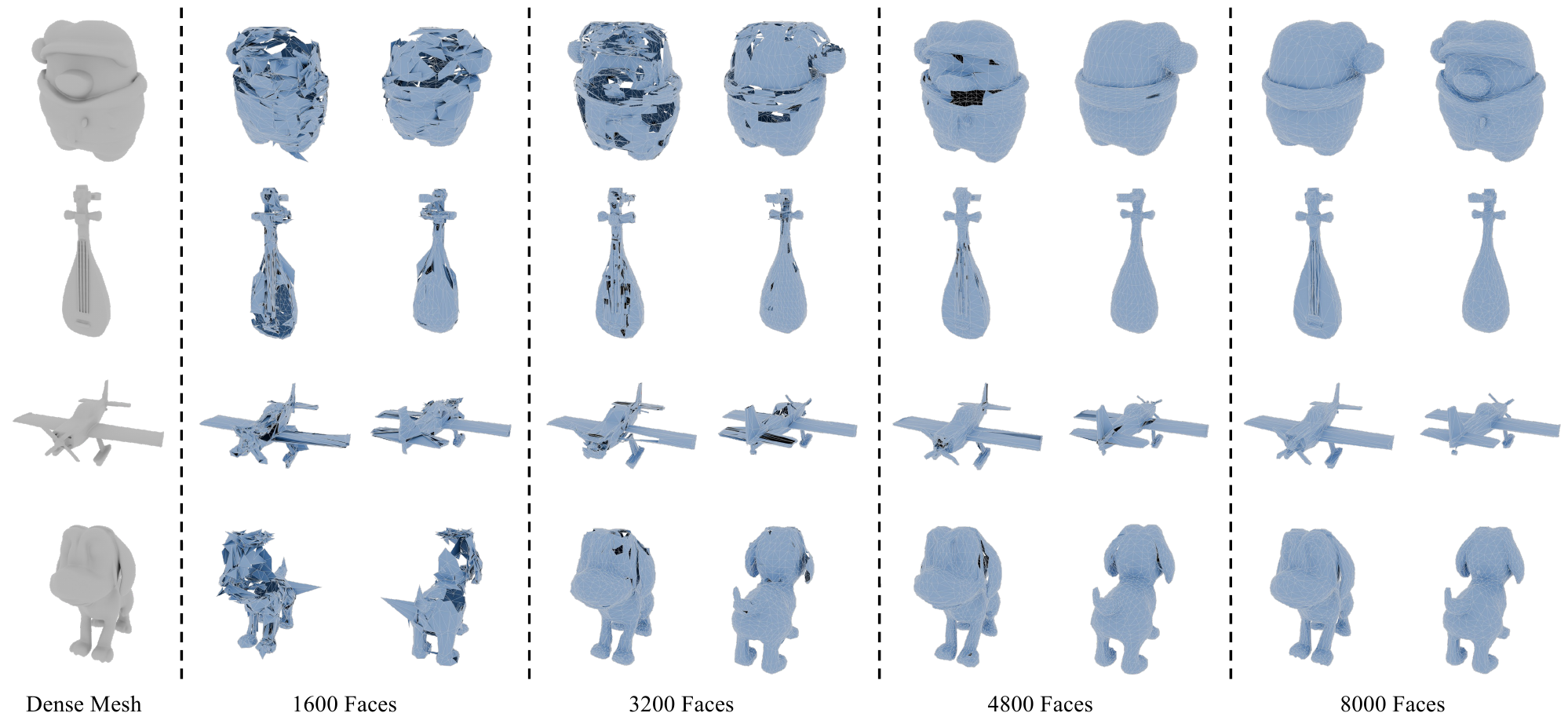}
\caption{\textbf{As the maximum face of the training meshes increases, the performance of mesh generation is significantly boosted. } 
It shows that scaling the mesh data with more faces is crucial to ensure the mesh generation performance. 
}
\label{fig: face_ablation}
\vspace{-4mm}
\end{figure*}

\subsection{Ablation Studies}

\paragraph{Select block size and offset size for BPT.}
The block size $|B|$ and offset size $|O|$ are crucial hyper-parameters for BPT.
We conduct the ablation experiments under the fixed mesh quantization resolution (i.e., $|B|\cdot|O|=128$).
With smaller block sizes, BPT's compression ratio is higher, thus making it better for mesh modeling. However, with too small block sizes (i.e., too large offset sizes), the vocabulary becomes unaffordable for prediction, thus reducing the generation performance.
As shown in Table \ref{tab: ablation}, the block size $8$ and offset size $16$ achieve the best generation performance.

\paragraph{Comparison with truncated training.}

There are also some engineering tricks to improve the utilization of training meshes. 
A typical solution is to train on truncated mesh sequences and inference with sliding window attention. 
We conduct the ablation study on truncated meshes with the Transformer context window of 4k and 8k, as shown in Figure \ref{fig: sw_ablation}. However, our experiments show that such truncation would reduce the robustness of mesh generation and make it easier to produce incomplete meshes. 

\vspace{-3mm}
\paragraph{Face matters for mesh generation.}

As the number of faces increases, the training meshes contain more details, improving the performance and robustness of mesh generative models.
As shown in Figure \ref{fig: data}(b) and Figure \ref{fig: face_ablation}, we conduct the ablation study with different training data, with the maximum faces of 1600, 3200, 4800, and 8000 separately. 
The experiment results show that scaling the mesh data with more faces is crucial to ensure the generation performance.

\section{Related Works}

\paragraph{3D generation with neural representation.}
Most previous attempts learn 3D shapes with various representations, e.g., SDF grids \citep{cheng2023sdfusion,chou2023diffusion,shim2023diffusion,zheng2023locally} and neural fields \citep{gupta3DGenTriplaneLatent2023,jun2023shap,muller2023diffrf,wang2023rodin,zhang20233dshape2vecset,liuMeshDiffusionScorebasedGenerative2023,lyu2023controllable}.
To improve the generalization ability, researchers start to leverage pre-trained 2D diffusion models \citep{rombachHighResolutionImageSynthesis2022,saharia2022photorealistic,liuZero1to3ZeroshotOne2023} with score distillation loss \citep{pooleDreamFusionTextto3DUsing2023,linMagic3DHighResolutionTextto3D2023,wangProlificDreamerHighFidelityDiverse2023} in a per-shape optimization manner.
Multi-view diffusion models \citep{shi2023mvdream,weng2023consistent123,zheng2023free3d,shi2023zero123++,chen2024v3d,voleti2024sv3d} are used to enhance the quality further and alleviate the Janus problem.
Recently, Large Reconstruction Models (LRM) \citep{hong2023lrm,li2023instant3d,xu2023dmv3d,wang2024crm,xu2024grm,tang2024lgm,xu2024instantmesh} train the Transformer backbone on large scale dataset \citep{deitke2023objaverse} to effectively generates generic neural 3D representation and shows the great performance of scaling.
However, these neural 3D shape generation methods require post-conversion \citep{lorensen1998marching,shen2021deep} for downstream applications, which is non-trivial and easy to produce dense and over-smooth meshes.

\vspace{-3mm}
\paragraph{Native mesh generation.}
\vspace{-2mm}
Compared with the well-developed generative models of neural shape representations, mesh generative models remain under-explored.
Some pioneering works try to tackle this problem by formulating the mesh representation as surface patches \citep{groueix2018papier}, deformed ellipsoids \citep{wang2018pixel2mesh}, mesh graphs \citep{dai2019scan2mesh}, and binary space partitioning \citep{chen2020bsp}.
PolyGen \citep{nash2020polygen}, Polydiff \citep{alliegro2023polydiff}, and MeshGPT \citep{siddiqui2023meshgpt} construct the promising generative models, but their performance is limited by the single-category datasets (e.g., ShapeNet).
Recent researches \citep{chen2024meshxl,chen2024meshanything,weng2024pivotmesh} 
build more generic generative models for native mesh generation within large-scale datasets.
Our research, along with several concurrent works \citep{chen2024meshanythingv2,tang2024edgerunner}, is designed to optimize the mesh representation, thus improving the model's scalability and robustness.

\section{Conclusion}

In this paper, we introduce Blocked and Patchified Tokenization (BPT), a fundamental improvement for mesh tokenization that reduces sequence length by approximately 75\%. 
Such compression efficiency utilizes meshes with over 8k faces, allowing us to scale the training data and enhance generation performance. 
By leveraging BPT, we have built a foundational mesh generative model conditioned on point clouds and images.

\vspace{-4mm}
\paragraph{Limitations and future work.}
The current number of parameters (500M) is still insufficient, and we will conduct additional scaling experiments with BPT.
Furthermore, we can explore other promising architectures for sequence modeling to better utilize the inductive bias of meshes.

{
    \small
    \bibliographystyle{ieeenat_fullname}
    \bibliography{main}
}

\clearpage

\appendix

\setcounter{page}{1}
\maketitlesupplementary

\section{Data Curation}

\begin{figure}[htbp]
\centering
\includegraphics[width=\linewidth]{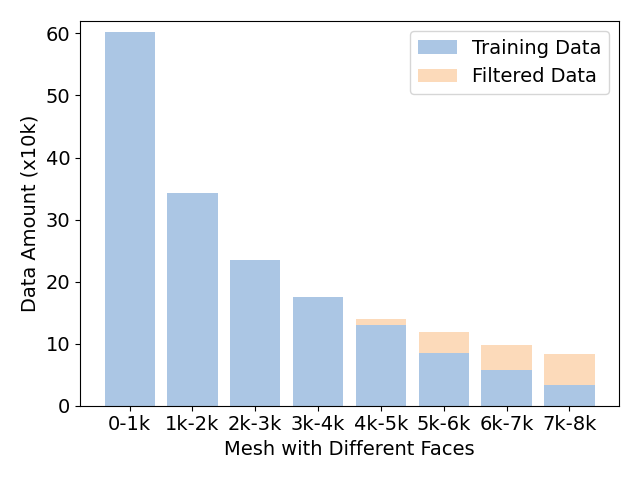}
\caption{\textbf{Effective utilization of training data.} 
Almost all meshes under 5k faces are used, and around 58\% of meshes with more than 5k faces are utilized further.
}
\label{fig: suppl-data}
\end{figure}

\paragraph{Effective data filtering.}
For meshes with the same faces, their tokenized sequence length may differ due to the patch aggregation and block compression of BPT.
We design an effective data-filtering strategy to maximize the utilization of our training data.
Specifically, we filter meshes with their sequence length lower than the context window of the Transformer (i.e., 9600).
Figure \ref{fig: suppl-data} shows that almost all meshes under 5k faces are used, and around 58\% of meshes with more than 5k faces are further utilized.
This strategy allows the utilization of some complicated meshes and improves the model's robustness and performance.

\paragraph{Two-stage training.}

Objaverse-xl contains many low-poly data with simple geometries, such as CAD meshes.
In the initial stage of training, the model may benefit from these meshes to learn the geometry prior. 
However, their topology is typically different from human-crafted meshes and may prevent the model from learning the delicate topology. 
Therefore, we leverage a two-stage training strategy to trade off the generalizability and topology quality. 
The model is first pretreated on the large-scale data with around 1.5M meshes and then further fine-tuned on 0.3M high-quality meshes without simple geometry.

\section{Model Architecture}

\begin{figure*}[!ht]
\centering
\includegraphics[width=0.8\textwidth]{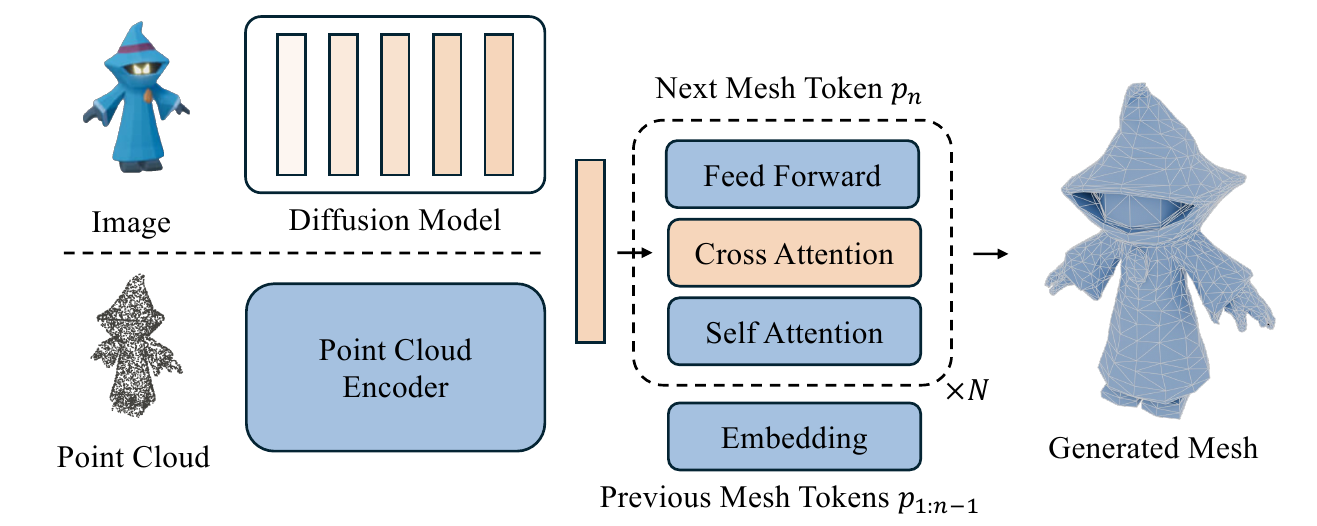}
\caption{\textbf{Model architecture for conditional mesh generation.} 
First, we leverage an auto-regressive transformer to generate meshes conditioned on point-cloud features via cross-attention layers.
Next, we train an additional diffusion model to generate point-cloud features based on images, enabling image-to-mesh generation.
}
\label{fig: suppl-arch}
\end{figure*}

As shown in Figure \ref{fig: suppl-arch}, the overall architecture of our model follows Michelangelo \cite{zhao2024michelangelo}.
As shown in Figure \ref{fig: suppl-arch}, we first train an auto-regressive transformer to generate meshes conditioned on point-cloud features extracted from the point-cloud encoder.
Then, we train an additional diffusion model to generate point-cloud features conditioned on images.

\section{Additional Results}

\paragraph{Comparison with remesh.}

\begin{figure*}[!ht]
\centering
\includegraphics[width=\textwidth]{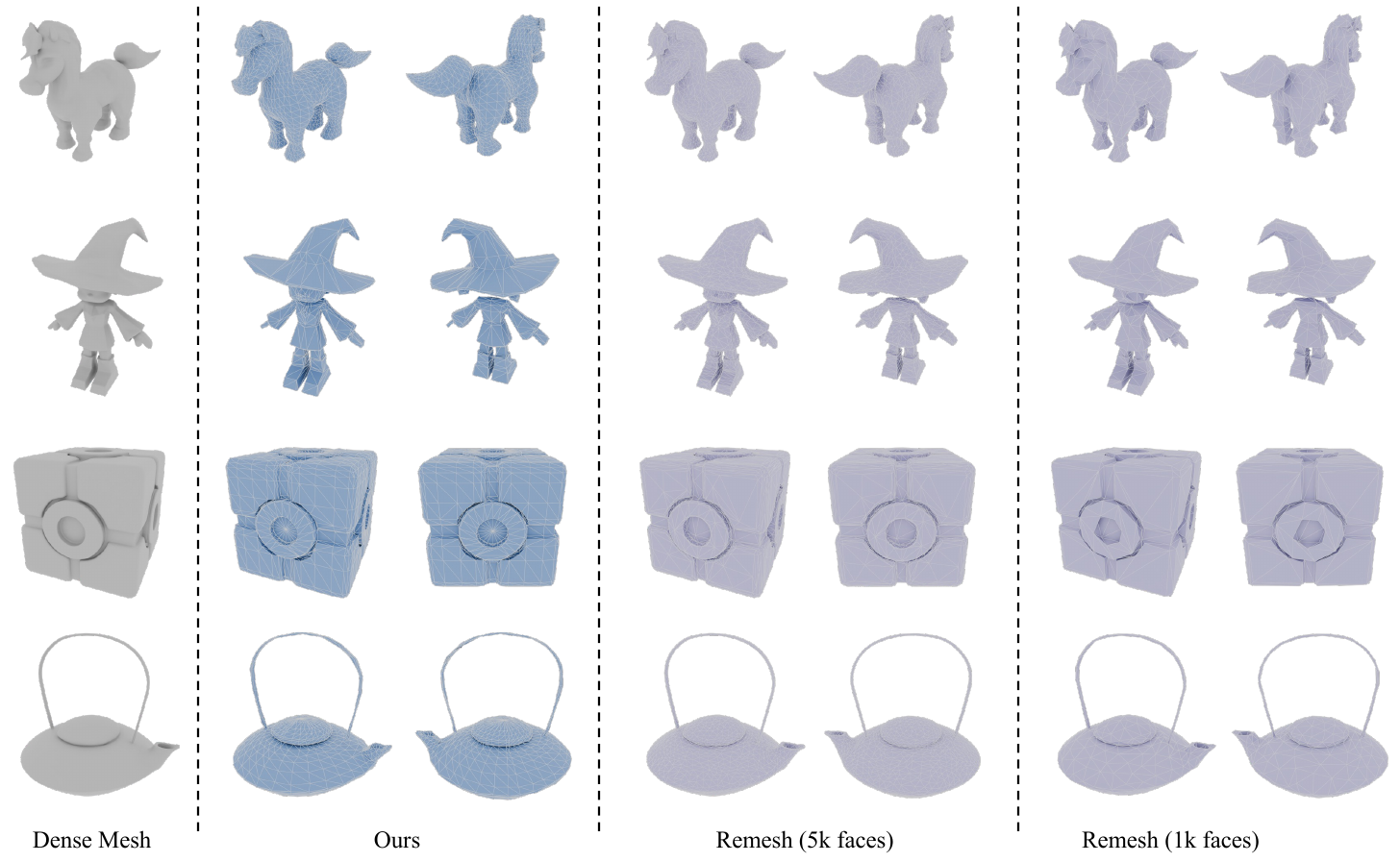}
\caption{\textbf{Comparison with remeshing.} 
Our method can generate appropriate topology from dense meshes, while remesh algorithms fundamentally fail to capture models' geometry and produce poor topology.
}
\label{fig: suppl-remesh}
\end{figure*}

Compared with remeshing algorithms, our method can generate appropriate topology from dense meshes, while remesh algorithms fundamentally fail to capture models' geometry and produce poor topology.
As shown in Figure \ref{fig: suppl-remesh}, the meshes generated by our model are at the product-ready level.

\end{document}